\documentclass[fleqn,twoside]{article}
\usepackage{espcrc2}

\usepackage{graphicx}
\usepackage[figuresright]{rotating}

\newcommand{\AmS}{{\protect\the\textfont2
  A\kern-.1667em\lower.5ex\hbox{M}\kern-.125emS}}

\title{Low-dimensional long-range topological structure in the QCD vacuum
        \thanks{Talk presented by I.~Horv\'ath, at Lattice 2003 Symposium, 
        Tsukuba, Japan.}
      }

\author{I.~Horv\'ath\address[UK]{Department of Physics \& Astronomy, 
        University of Kentucky, Lexington, KY 40506, USA},
        S.J.~Dong\addressmark[UK],
        T.~Draper\addressmark[UK],
        F.X.~Lee\address{Center for Nuclear Studies, 
        George Washington University, Washington, DC 20052, USA}\address[JL]
        {Jefferson Lab, 12000 Jefferson Avenue, Newport News, VA 23606, USA},
        K.F.~Liu\addressmark[UK],
        N.~Mathur\addressmark[UK],
        J.B.~Zhang\address{CSSM and Dept. of Physics and Math. Physics, 
        University of Adelaide, Adelaide, SA 5005, Australia},
        H.B.~Thacker\address[UVA]{Department of Physics, University of Virginia,
        Charlottesville, VA 22901, USA}}
       
\begin{document}

\begin{abstract}
Lattice topological charge associated with Ginsparg-Wilson fermions exhibits generic 
topological stability over quantum ensemble of configurations contributing to the QCD 
path integral. Moreover, the underlying chiral symmetry leads to the suppression of 
ultraviolet noise in the associated topological charge densities ({\em``chiral 
smoothing''}). 
This provides a solid foundation for the direct study of the role of topological
charge fluctuations in the physics of QCD vacuum. Using these tools it was recently 
demonstrated that: (a) there is a well-defined space-time structure (order) in topological 
charge density (defined through overlap fermions) for typical configurations contributing 
to QCD path integral; (b) this fundamental structure is low-dimensional, exhibiting 
sign-coherent behavior on subsets of dimension less than four and not less than one; 
(c) the structure has a long-range global character (spreading over maximal space-time 
distances) and is built around the locally one-dimensional network of strong fields 
(skeleton). In this talk we elaborate on certain aspects and implications of these results.
%\vspace{1pc}
\vspace{-0.12in}
\end{abstract}

% typeset front matter (including abstract)
\maketitle
\input epsf

The aim of this talk is to emphasize three messages (given in the abstract) from 
Ref.~\cite{Hor03A}. To appreciate the relevance of these messages, it is useful to recall 
that while the topology of gauge fields is presumed to be associated with interesting 
non-perturbative physics in QCD, the progress in making a detailed connection has been
quite slow. One possible path toward the breakthrough uses the lattice, where one has 
direct access to typical configurations contributing to the regularized QCD path integral. 
In principle then, one could study the space-time structure of topological charge density 
for typical configurations, and identify the characteristics of this structure that have 
direct bearing on the physical phenomena in question. However, there used to be several 
difficult obstacles as well as questions of principle preventing the realization of such 
a direct program.
  
(i) Thinking in terms of topology implies the availability of quantities that are stable 
under continuous deformations. The simplest topological field in the continuum is the 
topological charge density (TChD), whose associated global charge is stable under the local 
changes of the gauge field ($\delta Q/\delta A_\mu(x)=0$) for sufficiently smooth fields. 
The basic problem on the lattice is that it is non-trivial to define a local topological 
field. In particular, naive discretizations of $F\tilde{F}(x)$ are not stable at all (see 
however Ref.~\cite{Geom}).

(ii) Even if a lattice topological field, stable over some class of lattice gauge fields 
is available, there still might be a deep problem. The point is that we are interested 
in the {\em quantum} theory defined by the path integral. To study the relevance topology 
in this setting, one should thus ensure that topological stability is guaranteed for typical 
configurations (not necessarily smooth) from the quantum QCD ensemble. 

(iii) Even with points (i) and (ii) fulfilled, we might not gain much since it is possible 
that there is no well-defined space-time structure in TChD. Indeed, the mere existence 
of the structure at the configuration level is a subject of debate. While the action 
favours infrared fluctuations, there are many more possibilities for ultraviolet ones. 
In fact, for the commonly used lattice gauge actions there is an extensive experience 
indicating that when naive TChD operators are used, the ultraviolet noise dominates, and 
there is no detectable space-time order at all. However, without the order, we can not 
learn anything about the vacuum, and the approach outlined above would fail. This 
is the problem of ultraviolet dominance or the ``entropy'' problem.

The point (i) is elegantly resolved by using the TChD associated with GW fermions 
(see also Ref.~\cite{Geom}). Indeed, for lattice Dirac operator satisfying 
$\gamma_5 D \gamma_5 = D^{\dagger}$ and $\{D,\gamma_5\} = D \gamma_5 D$, the pseudoscalar 
operator $q_x \,=\, -\mbox{\rm tr} \,\gamma_5 \, (1 - \frac{1}{2}D_{x,x})$ represents
a well-defined local topological charge density~\cite{Has98A,Lus99A}, whose integral
is stable under generic local changes of the gauge field~\cite{Nar94A}. Morover, this
$q_x$ is generically stable for all backgrounds for which $D$ itself is well-defined! 
The accumulated experience with the overlap Dirac operator~\cite{Neu98A} (used here) 
indicates that this includes typical configurations from currently used QCD ensembles, 
thus fulfilling (ii).
  
Conceptually the most important result of Ref.~\cite{Hor03A} is that the point (iii) is 
also fulfilled if $q_x$ associated with overlap fermions is used, i.e. the conclusion (a) 
of the abstract holds and the problem of ultraviolet dominance is solved. We emphasize that 
this is a qualitatively new result since, contrary to previous studies of structure which 
always involved some particular processing of gauge configurations, our approach is 
completely unbiased and gauge invariant: we just use the appropriate local operator 
to measure TChD. Also note that the structure uncovered this way is the {\em fundamental 
structure} in the sense that (while suppressing ultraviolet noise) it includes 
the fluctuations at all scales up to the lattice cutoff. It can thus have manifestations 
reflecting both {\em short and long-distance} physics. 

An important clue in finding this structure is the old observation~\cite{SeSt}, that 
in the continuum $\langle q(x) q(0)\rangle \le 0,\, |x|>0$. As pointed out in 
Ref.~\cite{Hor02B}, this suggests that the fundamental structure cannot involve 
topological charge concentrated mainly in sign-coherent 4-d lumps. (The results 
of~\cite{Hor03A} 
not only demonstrate this but also suggest that in QCD the 4-d sign-coherent structure 
does not occur at all.) Given that, the starting idea of Ref.~\cite{Hor03A} is that 
the ordered structure could manifest itself by the enhanced sign-coherence 
present on {\em low-dimensional} subsets of 4-d Euclidean space. Indeed, the negativity 
of the correlator could then be satisfied in an ordered manner by embedding structures 
with alternating sign in 4-d space. The existence of such order has in turn been 
demonstrated~\cite{Hor03A}. 

%\begin{figure}
%\vspace*{-1.2in}
%\centerline{\includegraphics[width=8cm]{q_12_02_m.eps}}
%\vspace*{-1.2in}
%\label{fig:20}
%\end{figure}

%\vspace*{0.05in}
%\epsfxsize 0.99\hsize
%%\centerline{\epsffile {q_12_02_m.eps}}
%\centerline{\epsffile {test.eps}}
%\vspace*{0.05in}

%\vspace*{-1.2in}
%\epsfxsize 1.25\hsize
%\centerline{\epsffile {rg2_2.eps}}
%\vspace*{-1.2in}

\begin{figure}
\vspace*{-1.33in}
\epsfxsize 1.26\hsize
\centerline{\epsffile {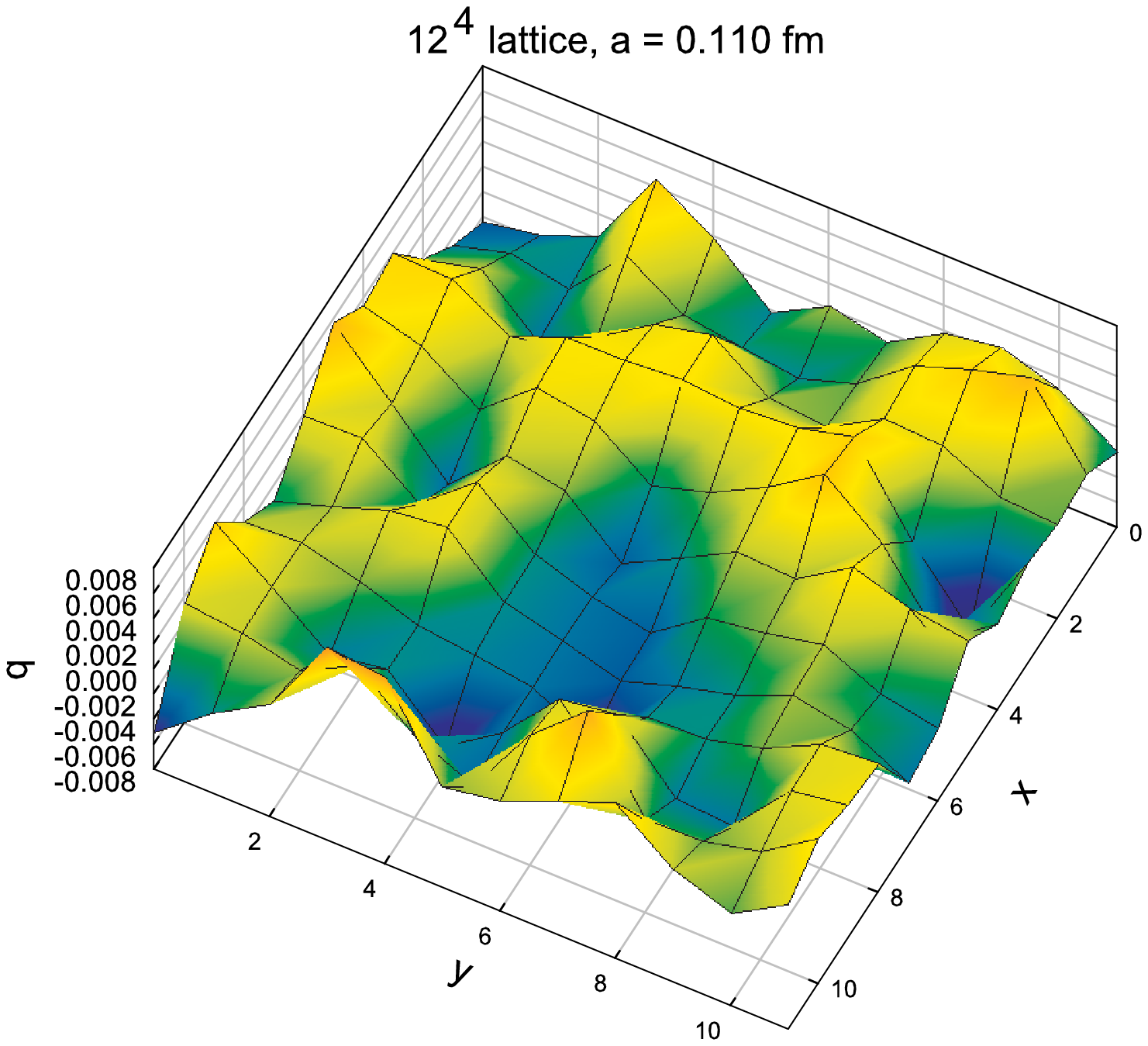}}
\vspace*{-1.20in}
\end{figure}

Using sign-coherence as a basic guide, it was found that about 80\% of space-time is 
covered by two connected folded ``sheets'' (of opposite sign) built from 3-d elementary 
coherent cubes. The sheets thus essentially fill the space-time. To get the sense of how 
this looks, we plot above $q_x$ on a (generic) 2-d plane of a $12^4$ configuration with
Wilson gauge action ($\beta=5.91$). The enhancement of the structure is obvious with 
linear ``ridges'' running across the whole system. Also, the low-dimensionality is 
nicely reflected in this generic picture. Indeed, if $\Omega$ is the base manifold and 
$\Omega_1,\Omega_2 \subset \Omega$, then 
$\dim ( \Omega_1 \cap \Omega_2 ) = \dim (\Omega_1) + \dim(\Omega_2) -  \dim(\Omega)$
generically. Identifying $\Omega$ with the 4-d space-time torus, $\Omega_1$ with the 2-d 
torus of the section plane, and $\Omega_2$  with the manifold of the structure, we get
\begin{equation}
   \dim ( \Omega_1 \cap \Omega_2 ) \,=\, \dim(\Omega_2) \,- \, 2
\end{equation}
Consequently, if $\Omega_2$ is a 3-d hypersurface, we should generically see extended 
1-d regions of sign-coherence. This is indeed what is observed. Thus, at the lattice 
cutoff of about 2 GeV the sheets behave as 3-d hypersurfaces. While the precise local 
dimension of the maximal sign-coherent regions in the continuum limit remains unknown, 
the available data allow us to conclude part (b) of the abstract~\cite{Hor03A}.

Apart from low-dimensional nature, the striking property of the TChD structure is its
long-range and global character. Indeed, the sheet appears to behave as one whole and 
due to connectedness and the space-filling feature, one can reach maximal possible 
distances on the paths within the same sheet. Moreover, this remains true even if one 
considers only regions occupied by strong fields. In particular, there is a well-defined 
minimal fraction of space-time ($\approx \!18\!$ \%) containing the most intense TChD 
while still exhibiting this {\em super-long-distance} (SLD) {\em property} \cite{Hor03A}. 
We refer to the substructure defined this way as a {\em skeleton} since the rest of the 
sign-coherent sheet is built around it. An intriguing property of the skeleton is that 
it is locally 1-dimensional~\cite{Hor03A}, and thus can be viewed as a network of 
world-lines for point-like objects. 

The above properties of the fundamental structure suggest the following two propositions
deserving further examination:
(A) The propagation of light quarks and pions is facilitated by the long-range (SLD) skeleton. 
One should thus try to relate the microscopic understanding of spontaneous chiral symmetry
breaking (SChSB) to the properties of this locally 1-dimensional structure. An interesting 
clue in making this connection is the fact that the motion of quarks restricted 
to 1-dimensional manifold naturally leads to non-zero density of eigenmodes near zero, 
and hence SChSB (see e.g. some related discussion \cite{Tik87}). 
(B) Global topological stability in quantum QCD configurations cannot be understood in terms 
of stability of its subsets, i.e. the global charge will not result as a sum of contributions 
from well-defined separate pieces. The global stability can only be understood if one considers 
the structure (or its SLD subset such as skeleton) as a whole. This can be considered a 
generalization of the result demonstrated in Refs.~\cite{Hor02B,Hor02A}, i.e. that topological 
charge does not appear in quantized lumps of unit topological charge (even at low energy). 

Finally, we comment on the mechanism leading to the resolution of the entropy problem (iii). 
The new element brought in by the imposition of chiral symmetry is that the GW operators 
are non-ultralocal, i.e. there is nonzero coupling among variables at arbitrarily large 
space-time distances. This has been proved for fermionic degrees of freedom \cite{Nultr}, but 
it is expected to hold also in terms of gauge variables. The implied presence of arbitrarily 
extended loops (albeit with small coupling) in the definition of $q_x$ leads to taming of 
ultraviolet noise at the scale of the cutoff to a certain degree ({\em ``chiral smoothing''}
\cite{Hor02A}). Moreover, the availability of these loops can help to uncover the long-range 
coherence (masked by ultraviolet noise) on extended low-dimensional manifolds since the loops 
contained inside these manifolds will contribute coherently. The results of \cite{Hor03A} 
imply that such coherence exists, and is indeed exposed by chiral smoothing.

\end{document}